\documentclass[a4paper]{jpconf}% no class options needed by now
\usepackage[english]{babel}
\usepackage{graphicx}
\begin{document}
\title{Status and prospects of searches for neutrinoless double beta decay}
% \subtitle{subtitle}
\author{Bernhard Schwingenheuer}
\address{Max-Planck-Institut f\"ur Kernphysik, 69117 Heidelberg, Germany}
\begin{abstract}
The simultaneous beta decay of two neutrons in a nucleus
without the emission of neutrinos 
(called neutrinoless double beta decay) is a lepton number violating 
process which is not allowed in the Standard Model of
particle physics.
More than a dozen experiments using different candidate
isotopes and a  variety of detection techniques are searching for this decay.
Some (EXO-200, Kamland-Zen, GERDA) started to take data
recently. EXO and Kamland-Zen have 
 reported  first limits of the
half life $T_{1/2}^{0\nu}$ for  $^{136}$Xe.
After a decade of little progress in this field, many new
results will soon scrutinize the claim from
part of the Heidelberg-Moscow collaboration to have observed this
decay.
The sensitivities of the different proposals are reviewed.
\end{abstract}
%\maketitle
% \noindent

\section{Introduction}

For 35 isotopes,  $\beta$ decay is energetically
forbidden but double beta decay ($2\nu\beta\beta$) is allowed
\begin{equation}
 (A,Z) \rightarrow (A,Z+2) + 2e^- + 2\bar{\nu}_e.
\end{equation}
This process has been observed directly for 11 isotopes 
with half lives between $7\cdot 10^{18}$~yr and
$2\cdot 10^{21}$~yr \cite{barabash2nu,exo,kamland}.

Since neutrinos have no electric charge, there is no
known symmetry which forbids that 
they  mix with their anti-particles.
As a consequence
double beta decay without neutrino emission (0$\nu\beta\beta$) 
might occur as well which is
predicted by almost all extensions of the Standard Model of
particle physics (see e.g.~\cite{rodejohann,ejiri}). 
Its observation would imply that lepton number is violated since
only electrons are emitted.

The experimental signature of $0\nu\beta\beta$ is a line at 
the $Q_{\beta\beta}$ value of the decay if the sum
of the electron energies is histogrammed.

Part of the Heidelberg-Moscow collaboration
claims to have evidence for this line for
$^{76}$Ge with $T_{1/2}^{0\nu} = (1.19_{-0.23}^{+0.37}) \cdot 10^{25}$~yr\cite{klapdorplb}
at a 4$\sigma$ level. In a more sophisticated analysis a significance
beyond $6\sigma$ was calculated \cite{klapdor2}.
The current experiments using $^{136}$Xe (EXO-200
and Kamland-Zen) and $^{76}$Ge (GERDA) will scrutinize
this result within the next 12 months.

In total more than a dozen large scale
experimental programs are suggested or under
construction to search for $0\nu\beta\beta$. These programs are
compared in this article and also the status of theoretical
matrix element calculations is discussed.
The latter are needed to convert an experimental measurement
or limit on $T_{1/2}^{0\nu}$ to a particle physics parameter. For general
reviews the reader is referred to the literature 
\cite{engel,rodejohann,gomez,ejiri}.

There are also other related processes like double  positron decay
or double electron capture. 
While $0\nu\beta\beta$ is already a suppressed process,
the other decays are expected to
be even rarer unless there is some resonance enhancement
\cite{suhonen1,tgv,danevich,ejiri,blaum}. 
In this article only $0\nu\beta\beta$ decay searches
are discussed.

\section{Motivation for $0\nu\beta\beta$}

Since neutrinos have mass but no electric charge,
there is no known symmetry which forbids 
additional terms in the effective Lagrangian besides the 
standard Dirac mass term $m_D$ \cite{engel,rodejohann,ejiri,bilenky}:
\begin{eqnarray}
 L_{\rm Yuk} & = & m_D \overline{\nu_L} \nu_R +  \frac{1}{2}
                m_L \overline{\nu_L}  (\nu_L)^c
              + \frac{1}{2} m_R \overline{(\nu_R)^c} \nu_R  + h.c. \\
            & = & \frac{1}{2}
  ( \overline{\nu_L},\,\, \overline{(\nu_R)^c} )
 \left( \begin{array}{cc} m_L & m_D \\ m_D & m_R \end{array} \right)
 \left( \begin{array}{c} (\nu_L)^c \\ \nu_R \end{array} \right) + h.c. 
\end{eqnarray}
The subscript $L$ stands 
for the left-handed chiral field $\nu_L = \frac{1}{2}(1-\gamma_5)\, \,\nu$ and
 $R$  for the right-handed projection 
 $\frac{1}{2}(1+\gamma_5)\,\, \nu$. The superscript $C$ 
 denotes charge conjugation, i.e.~$\nu^C$  stands for an (incoming) anti-neutrino.
The $m_R$ term describes therefore an  incoming neutrino $\nu_R$ 
and an outgoing
anti-neutrino $\overline{(\nu_R)^C}$, 
i.e.~this term violates lepton number  by 2 units.
The eigen states of the mass matrix are of the
form $(\nu+\nu^c)$. Consequently neutrinos are 
expected to be - in general - their
own anti-particles, i.e.~Majorana particles.

Neutrinos (or anti-neutrinos)
are produced in charged weak current reactions and - depending on
the charge of the associated lepton -  only one chiral 
projection couples (``V-A'' current of weak interactions).
For example in $\beta$ decay $n\rightarrow p \, e^- \, \bar{\nu}_{e,R}$,
a right-handed anti-neutrino couples which can be
decomposed in the mass eigen states $\nu_i$ and helicity ($h=\vec{\sigma}\vec{p}$)
eigen states:
\begin{equation}
\label{eq:nur}
  \bar{\nu}_{e,R} = \bar{\nu}_e \frac{1}{2}(1+\gamma_5)
           = \sum\limits_{i=1}^{3} U_{ei} (\bar{\nu}_{i,h=+1} + 
                \frac{m_i}{E}\bar{\nu}_{i,h=-1})
\end{equation}
Here, $U$ is the PMNS mixing matrix \cite{pdg},
$m_i$ are the  mass eigen values, and
$E$ is the neutrino energy.

For  Dirac particles, only
detection reactions like $p\, \bar{\nu}_{e,R} \rightarrow n\, e^+$
are possible.
If, on the other hand, neutrinos are massive Majorana particles, then
the helicity suppressed component $\bar{\nu}_{i,h=-1}$  of $\bar{\nu}_{e,R}$
 can undergo the reaction 
 $n \, \nu_{e,L}  \rightarrow p \, e^-$. Here
\begin{equation}
\label{eq:nul} 
  {\nu_{e,L}} = \frac{1}{2}(1-\gamma_5) {\nu_e} 
           = \sum\limits_{i=1}^{3} U_{ei} ({\nu_{i,h=-1}} + 
                \frac{m_i}{E} {\nu_{i,h=+1}})
\end{equation}
Taking both processes involving neutrons together
we have $2\,n \rightarrow 2\, p + 2\, e^-$ (or better
$  (A,Z) \rightarrow (A,Z+2) + 2e^-  $)
 mediated 
by a massive Majorana neutrino with an effective coupling strength
which is called the Majorana mass:
\begin{equation}
\label{eq:mbb}
  m_{\beta\beta}=|\sum\limits_{i=1}^3 U_{ei}^2\cdot m_i|
\end{equation}
The  helicity suppression $(m_i/E)^2$, which is e.g.~$10^{-14}$
for a neutrino mass of 0.1~eV and a neutrino energy of 1~MeV, 
is compensated by the large number of nuclei per mole.

Schechter and Valle showed that the observation of $0\nu\beta\beta$
ensures that neutrinos have a Majorana component \cite{schechter}.
Recently it was pointed out that the ``guaranteed'' Majorana mass
through radiative corrections
is however only in the range of  $10^{-24}$~eV \cite{duerr}, i.e.~negligible
compared to the mass scales of neutrino oscillations \cite{pdg}.
Consequently, if $0\nu\beta\beta$ is observed, other mechanisms
like the exchange of supersymmetric particles or heavy Majorana neutrinos
 might be the dominating process  and the known
neutrino could even be (effectively) a  Dirac particle. 
While for the initially motivated light neutrino exchange the coupling strength
is proportional to $m_i$ (Eq.~\ref{eq:mbb}),   the one for processes with
heavy fermion exchange is proportional to $1/M_F$ with $M_F$ being the
mass of the exchange particle (provided that the helicity
of the two leptonic currents are the same 
and $M_F^2 > q^2$).\footnote{To be more precise: the amplitude has
dimension mass$^{-5} = M_B^{-4}\cdot M_F^{-1}$ with $M_B$ being the mass
of a scalar or vector exchange particle like the $W$ boson.}
In such processes, lepton number violation or lepton flavor violation
can be accessible with accelerator experiments as well \cite{lhcb}.
If lower limits on $M_F$ from e.g.~LHC are higher than
typically 10~TeV, then the contributions to $0\nu\beta\beta$
become smaller than the ones expected from light neutrino
exchange with a mass of 0.05~eV \cite{rodejohann}.
The argument can be turned around:  $0\nu\beta\beta$ will
provide also information on TeV scale physics.

The exchange of light neutrinos is discussed predominantly.
Its strength depends on $U$ which
can be parameterized by 3 rotation angles and 1 phase
($\theta_{12}$, $\theta_{13}$, $\theta_{23}$, and $\delta$, all
measurable by neutrino oscillation experiments), and 2 additional phases
($\alpha_{21}$ and $\alpha_{31}$). The latter are called Majorana
phases and
influence processes like $0\nu\beta\beta$.
Oscillation experiments measure difference of squared masses
($\Delta m_{21}^2=m_2^2-m_1^2=(7.58^{+0.22}_{-0.26})\cdot10^{-5}$~eV$^2$ and
$|\Delta m_{31}|^2=|m_3^2-m_1^2|=(2.35^{+0.12}_{-0.09})\cdot 10^{-3}$~eV$^2$ \cite{pdg}).
Knowledge on the absolute mass scale comes 
from e.g.~beta decay, $0\nu\beta\beta$  or cosmology.

One can estimate possible values of $m_{\beta\beta}$ as a function
of the lightest neutrino mass given the experimental knowledge
about $\Delta m_{ij}^2$ and the measured rotation angles and
allowing the Majorana phases to take any value.
Three cases can be discriminated \cite{vissani}.
\begin{itemize}
\item If the lightest mass  is larger than all $\Delta m_{ij}^2$,
      all $m_i$ are similar (degenerate masses).
\item If $m_1$ is the smallest mass (normal mass hierarchy) the 3 terms
      of $m_{\beta\beta}$ can cancel for $m_1 \leq \sqrt{\Delta m_{21}^2}$. 
      For $m_1\approx 0$ typical values are a few
      meV.
\item If $m_3$ is the smallest mass (inverted hierarchy, $\Delta m_{31}^2<0$)
       there exists for $m_{\beta\beta}$
      a lower bound of $(19^{+1.7}_{-1.5})$~meV 
       and an upper bound of 50~meV  \cite{bilenky}.
\end{itemize}
If additional sterile neutrinos exist, i.e.~if there are additional
terms in Eq.~\ref{eq:mbb}, then  even
for the inverted hierarchy no lower bound exists \cite{barry,giunti}.

In summary, $0\nu\beta\beta$ - together with other input from
e.g.~neutrino oscillation and LHC experiments - provides 
an important window to  extensions of the Standard Model
with lepton number violation.

\section{Experimental sensitivity on $T_{1/2}^{0\nu}$}

All (but one) searches have so far only observed  event
counts in the region of interest around $Q_{\beta\beta}$ which are consistent
with the expectation from background $\lambda_{\rm bkg}$. This
number
- if it scales with the detector mass $M$ - is given by
\begin{equation}
\label{eq:nbkg}
\lambda_{\rm bkg} = M \cdot t \cdot B \cdot \Delta E
\end{equation}
Here  $t$ is the
measurement time, $B$ is the so called background
index given typically in cnts/(keV$\cdot$kg$\cdot$yr),
and $\Delta E$ is the width of the search window which
depends on the experimental energy resolution.
Note that this equation is only an approximation.
Experiments normally take the (nonlinear)
shape of the background spectrum in a fit into account.

The non-observation is converted to an upper limit on
the number of signal events $\lambda_{\rm sig}$ which
is related to the half life $T_{1/2}^{0\nu}$ of
a given isotope $A$ by
\begin{equation}
\label{eq:nsig}
  \lambda_{\rm sig} = \ln 2 \cdot N_{\rm Avg} \cdot \epsilon \cdot
    \eta \cdot M \cdot t /(m_A \cdot T_{1/2}^{0\nu}).
\end{equation}
$N_{\rm Avg}$ is  the Avogadro constant, $\epsilon$ the
signal detection efficiency, $\eta$ the mass fraction
of the $0\nu\beta\beta$ isotope, and
$m_A$ the molar mass of the isotope.

\begin{table}
\caption{\label{tab:g0nu} List of phase space factor (Eq.~\ref{eq:nme11}),
$Q_{\beta\beta}$, natural abundance of $\beta\beta$ isotope,
half life of $2\nu\beta\beta$, and experiments  
for the most interesting $0\nu\beta\beta$
isotopes. Half lives for $2\nu\beta\beta$ 
are taken from \cite{barabash2nu,exo},
$Q_{\beta\beta}$ for $^{136}$Xe from \cite{redshaw}, and all
other numbers  from \cite{rodejohann}. The $G^{0\nu}$ values
have been scaled to the same nuclear radius and $g_A$ coupling.}
\begin{center}
\begin{tabular}[\columnwidth]{lccccl}\hline
 isotope  & $G^{0\nu}$    & $Q_{\beta\beta}$ & nat.~ab. & $T_{1/2}^{2\nu}$ & experiments \\
          & $[\frac{10^{-14}}{\rm yr}$] & [keV]        &  [\%]       & [10$^{20}$ y] &  \\ \hline
 $^{48}$Ca & 6.3  & 4273.7  & 0.187  & 0.44 & CANDLES \\
 $^{76}$Ge & 0.63 & 2039.1  & 7.8   & 15 & GERDA, Majorana Demonstr. \\
 $^{82}$Se & 2.7  & 2995.5  & 9.2   & 0.92 & SuperNEMO, Lucifer \\
 $^{100}$Mo & 4.4 & 3035.0  & 9.6   & 0.07 & MOON, AMoRe \\
 $^{116}$Cd & 4.6 & 2809.1  & 7.6   & 0.29 & Cobra \\
 $^{130}$Te & 4.1 & 2530.3  & 34.5  & 9.1 & CUORE \\
 $^{136}$Xe & 4.3 & 2457.8  & 8.9   & 21 & EXO, Next, Kamland-Zen \\
 $^{150}$Nd & 19.2 & 3367.3 & 5.6   & 0.08 &SNO+, DCBA/MTD \\ \hline
\end{tabular}
\end{center}
\end{table}

If $\lambda_{\rm bkg} < 1$ the experimental sensitivity
scales with $M\cdot t$ while for $\lambda_{\rm bkg} >> 1$
the e.g.~90\%\, C.L.~limit on the half life (assuming there
is no signal) is given by
\begin{equation}
\label{eq:comp}
         T_{1/2}^{0\nu}(90\% CL) > \frac{\ln 2}{1.64} \frac{N_{\rm Avg}}{m_A}\,
               \epsilon\cdot \eta \cdot\sqrt{\frac{M\cdot t}{B\cdot \Delta E}}. 
\end{equation}
If systematic errors become important e.g.~if the energy resolution
or the spectral background shape is not well known,
then the sensitivity is reduced.

\section{The nuclear matrix element}

$(T_{1/2}^{0\nu})^{-1}$ is the product of three factors:
a phase space factor $G^{0\nu}$, a nuclear matrix element $M^{0\nu}$ and a particle
physics factor.

If the exchange of light Majorana neutrinos is dominating,
$T_{1/2}^{0\nu}$ for a given isotope $A$ is  \cite{doi}
\begin{equation}
\label{eq:nme11}
  [T_{1/2}^{0\nu}(A)]^{-1} = 
        G^{0\nu}(Q_{\beta\beta},Z) \cdot |M^{0\nu}(A)|^2 \frac{m_{\beta\beta}^2}{m_e^2}
\end{equation}

Here, $m_e$ is the electron mass.
Values for $G^{0\nu}$ are listed in Tab.~\ref{tab:g0nu}. 
Obviously, $M^{0\nu}(A)$ is needed
to compare results from different isotopes or to extract
information about the particle physics parameter. 

\begin{figure}
\begin{minipage}[t]{0.48\textwidth}
\includegraphics[width=\textwidth]{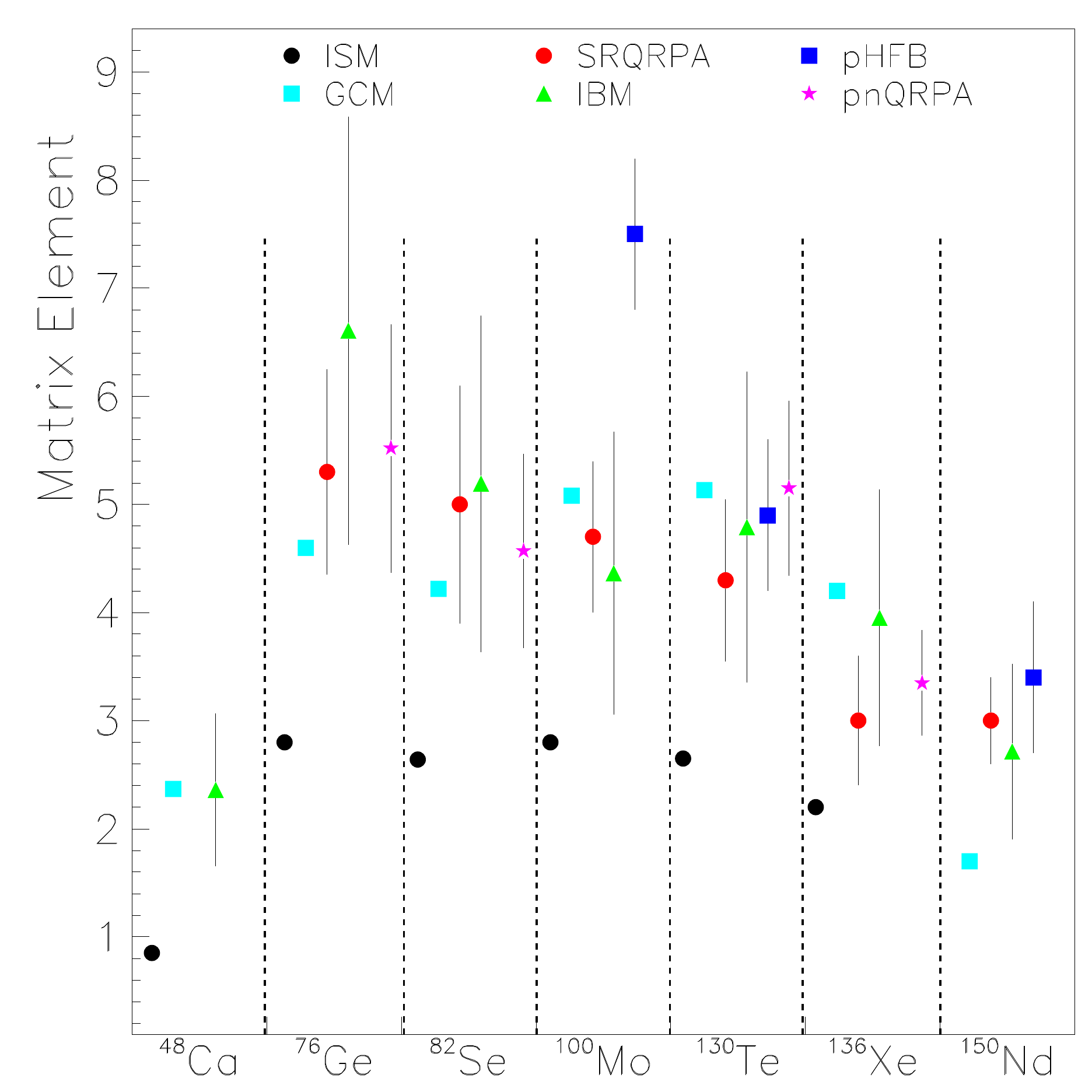}
\caption{\label{fig:nme} Nuclear matrix element calculations 
  for $0\nu\beta\beta$  for light neutrino exchange. 
  ISM = Interacting shell model \cite{nsm,nsm3}, 
  SRQRPA = self-consistent renormalized quasi-particle random 
  phase approximation \cite{tuebingen12},
   pnQRPA = proton-neutron quasi particle random phase 
   approximation \cite{suhonen2},
  GCM = generating coordinate method \cite{gcm}, IBM = interacting
  boson model \cite{ibm,ibm2,ibm3} (matrix elements are scaled by 1.18 to estimate the
  effect if the UCOM short range correlation instead of the Jastrow type would
  have been used \cite{gomez}, in \cite{ibm2} an error of 30\% is estimated), pHBF=
  projected Hartree-Fock-Bogoliubov model \cite{phfb}.}
\end{minipage}
\hspace{1mm}
\begin{minipage}[t]{0.48\textwidth}
 \includegraphics[width=\textwidth]{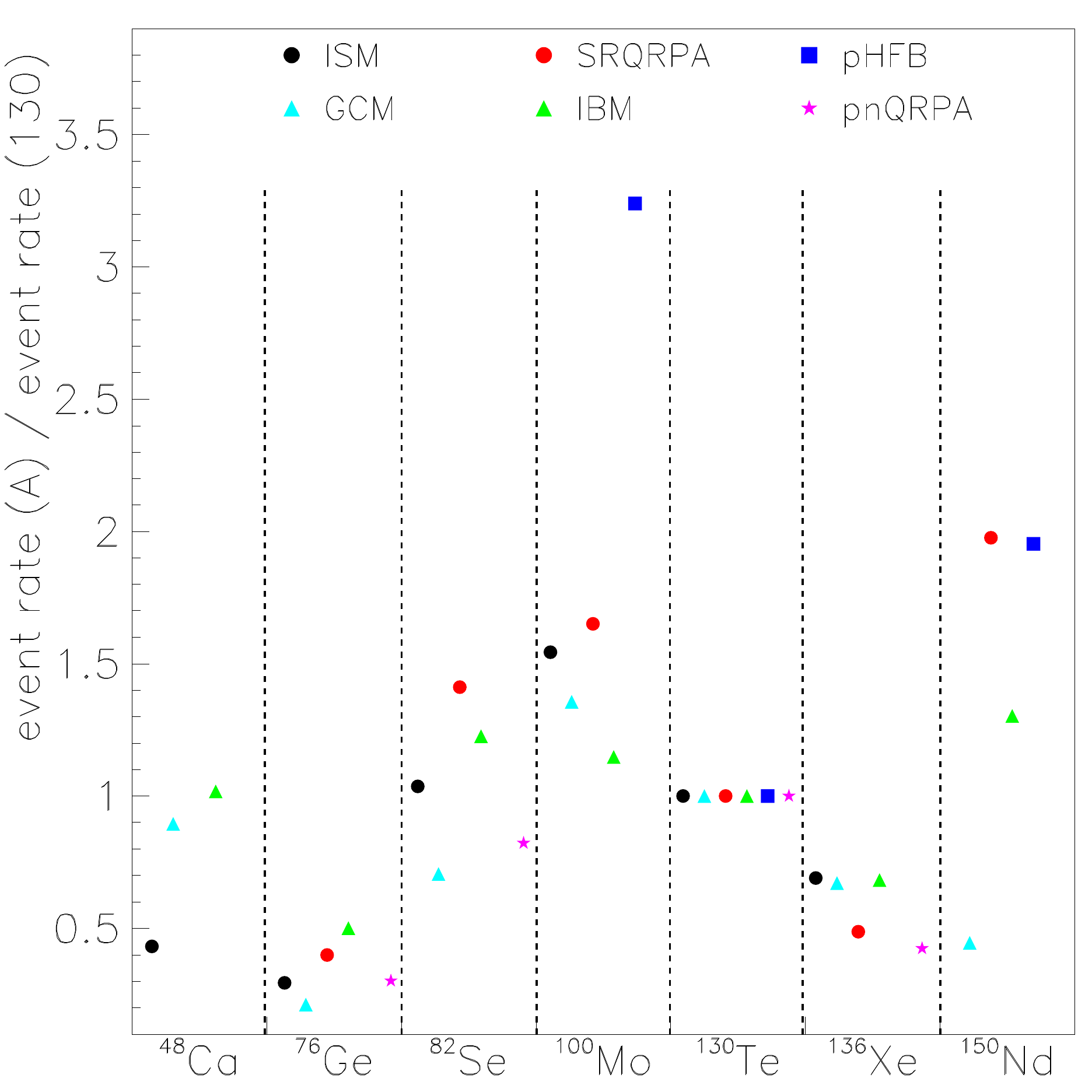}
\caption{\label{fig:nme2} 
  Ratio of expected $0\nu\beta\beta$ events per kg target mass
  for the different matrix element models normalized to $^{130}$Te.}
\end{minipage}
\end{figure}

The $M^{0\nu}$ calculations
are difficult and can only be done using approximations.
Traditionally, interacting shell model (ISM) \cite{nsm,nsm3}
 and quasi particle random phase 
approximation (QRPA) calculations have been performed 
 \cite{suhonen2,srqrpa2}.
Recently new approaches like the interacting boson model (IBM) \cite{ibm,ibm2,ibm3},
the generating coordinate model (GCM, also called
energy density functional EDF) \cite{gcm} and the projected 
Hartree-Fock-Bogoliubov (pHFB) method \cite{phfb} have been applied.

The results of these calculations are shown in Fig.~\ref{fig:nme}.
The following statements can be made concerning the status:
\begin{itemize}
\item $M^{0\nu}(A)$ varies slowly with $A$.
   This might be due to the
   fact that only neighboring neutrons in a nucleus contribute
   to the decay  \cite{nsm,srqrpa2}. 
\item For the ISM, all values are systematically lower than for other methods.
      Possible reasons for this effect are discussed in the literature
      \cite{nsm,escuderos}.
%\item The differences between the QRPA calculations of different
%      groups are now quite small.
\item For a given isotope the calculations spread by typically a factor of 2-3,
      i.e.~a factor of 4-9 for $T_{1/2}^{0\nu}$. 
      Some groups  estimate a possible range for  $M^{0\nu}$ 
      (see indicated errors in Fig.~\ref{fig:nme}).
      However it is unclear what
      to quote as confidence interval for theoretical calculations.
\item The role of short range correlations has been studied and the UCOM
      correction has emerged as favorable \cite{civitarese}. 
      Alternatively, a self 
      consistent implementation
      was first applied to SRQRPA \cite{srqrpa2}
      and later to other methods \cite{nsm2,phfb} and resulted in
      small changes.
\item Experimental input can cause a sizable shift of the result. For example
      charge exchange reaction measurements of $^{150}$Nd($^3$He,t)
      and  $^{150}$Sm(t,$^3$He) \cite{zegers}
      result in a quenching factor of 0.75 for the
      $g_A$ coupling and hence a reduction of the matrix element by 25\%
      for $^{150}$Nd \cite{srqrpa3}. In this calculation, deformation
      was treated for the first time in a QRPA calculation.

      For $^{76}$Ge and $^{76}$Se,  the proton and neutron valence orbital 
      occupancies have been measured \cite{schiffer,schiffer2}. 
      If the models are adjusted to reproduce
      these values, the ISM result increases by 15\% \cite{nsm2} 
      while the QRPA
      results are reduced by about 20\% \cite{simkovic,suhonen3}. 
      Hence the difference between ISM and QRPA  becomes  half as large.
\end{itemize}

In order to see whether some isotopes are better suited
for $0\nu\beta\beta$  searches from a theoretical  point of view,
the number of expected decays for a given exposure
(in units of kg$\cdot$yr) can be compared using
Eqs.~\ref{eq:nsig} and \ref{eq:nme11} without 
the factors $\epsilon$ and $\eta$. The $A$ dependent parameters
are then the phase space factor, the matrix element and
the molar mass. For the comparison it is sufficient to look at
the ratio of decay rates and in this case, some of the systematic effects
of the matrix element calculations cancel since there are
typically correlations among the isotopes for a given method.
\footnote{This can be seen for example from Tab.~1 of reference
\cite{tuebingen12}. For every isotope, 20 different calculations 
 are listed which vary by typically
a factor of 2 for a given $A$. If  20 ratios $M^{0\nu}(A)/M^{0\nu}(A')$
for two isotopes $A$ and $A'$ are calculated, the variation is reduced to 30\%.}
 Fig.~\ref{fig:nme2}
shows these ratios for the different models normalized to the
decay rate of $^{130}$Te. One sees that $^{76}$Ge is less favorable.
The expected decays per kg  vary between 20\% and 50\% of the
rate of $^{130}$Te. In other words: if all experimental parameters
(number of background events, efficiency, etc.)
were the same then one would need a factor of 2-5 more target mass
in a $^{76}$Ge experiment to have the same sensitivity.
In reality, all experimental parameters like the energy resolution
and the background have to considered as well.

\section{Current  experimental situation}

In the last decade mainly three experiments contributed
to  $0\nu\beta\beta$ searches. Heidelberg-Moscow was
for more than a decade the most sensitive one and
reported evidence for this decay. The two others
were Cuoricino and NEMO-3. Before, IGEX reported
with an exposure of 8.8~kg$\cdot$yr
a limit of $T_{1/2}^{0\nu}>1.57\cdot 10^{25}$~yr for $^{76}$Ge 
(90\%~C.L.)\cite{igex}.
Currently EXO-200, Kamland-Zen, and GERDA 
are taking data. All experiments are discussed in this section. 

\subsection{The Heidelberg-Moscow experiment}

The Heidelberg-Moscow experiment operated between 1990 and 2003
five germanium detectors
made out of isotopically enriched material ($\simeq$86\% $^{76}$Ge, 11~kg).
 The diodes were mounted in copper
cryostats with copper, lead, and polyethylene shielding.
The total exposure was 71.7~kg$\cdot$yr
and the average count rate in the interval 2-2.1~MeV was
about  0.17 cnts/(keV$\cdot$kg$\cdot$yr) (for the period 1995-2003). 
The energy resolution (full width at half maximum, FWHM)
was about 3.5~keV at $Q_{\beta\beta}$ which is the best value
of all $0\nu\beta\beta$ experiments.
Part of the collaboration finds evidence for a peak
at $Q_{\beta\beta}$ with $28.75\pm6.86$ events which converts
to  $T_{1/2}^{0\nu} = (1.19^{+0.37}_{-0.23})\cdot 10^{25}$~yr \cite{klapdorplb}.
Note that  only a statistical error is quoted.
Another study finds
that e.g.~extending the energy window used in the
data fit increases this background and hence decreases the 
signal count by up to 40\% (Tab.~3.8 and 4.6 of reference \cite{oleg}).

In a later publication \cite{klapdor2} the claim was strengthened by a pulse
shape analysis which preferentially selected $0\nu\beta\beta$ events
due to their localized energy deposition in  the detectors.
Backgrounds from gammas with multiple Compton scatterings
exhibit different pulse shapes.
The background is reduced to a surprisingly low level 
of $\approx$0.015~cnts/(keV$\cdot$kg$\cdot$yr). The final fit  reports
a yield of 11.32$\pm$1.75 signal events 
and the ratio $11.32/1.75 = 6.5$ is  called
the significance of the peak (Fig.~9b in \cite{klapdor2}).
The signal yield  was then converted
to $T_{1/2}^{0\nu}=(2.23^{+0.44}_{-0.31})\cdot 10^{25}$~yr.
There are several problems with this analysis.
\begin{itemize}
\item The fit error on the signal count is too small.
      The smallest 68\% Poisson credibility interval is between 8.1 and
      15.2 for a probability distribution which peaks at 11.3,
      i.e.~a factor of 2 larger than the quoted interval.
      Due to the existing (small) 
      background the $\pm 1\sigma$ interval should become
      even larger.
\item The probability that the background ($\simeq 2.2$ events
      in the central 3 keV of the peak) fluctuates to the observed number
      of 13 events or more is 5$\cdot 10^{-7}$ which converts
      to a significance of about 5$\sigma$. Systematic
      effects like the uncertainty of the background might
      reduce this value.
\item In the conversion to $T_{1/2}^{0\nu}$ using Eq.~(\ref{eq:nsig}), 
       an efficiency
      $\epsilon$ of 100\% is used although no value is  explicitly quoted.
      All but three events in the peak are part of an earlier selection
      (labelled ``HNR+NN'' in \cite{klapdor2}). For the latter the efficiency
      was 62\%  \cite{klapdornim}. Hence one expects also for this
      analysis a value much smaller than 100\%.
\end{itemize} 
The central $T_{1/2}^{0\nu}$ value and the errors are consequently not correct
in \cite{klapdor2}  and the significance is smaller than quoted although
still high. 
%Since important information (especially the
%efficiency) can not be extracted, the  $T_{1/2}^{0\nu}$
%result of this publication should
%not be used in the discussion any longer.

\subsection{The Cuoricino experiment}

Cuoricino \cite{cuoricino}
%e andreotti et al, Astropart. Phys 34 (2011)  822-831.
operated 62 TeO$_2$ crystals with a total mass of 40.7~kg
(11.3~kg of $^{130}$Te) between 2003 and 2008.
At a temperature of $\leq 10$~mK, the heat capacitance is very low
and an energy
deposition inside a crystal results in an increase
of typically 0.1~mK/MeV which is measured with Neutron Transmutation Doped
germanium thermistors. The latter have
a resistance of 100~M$\Omega$ and show a strong temperature dependence
which converts to a dependence of 3~M$\Omega$/MeV.
This bolometric technique has been proven
to work with a good resolution FWHM of typically 6-10~keV at 2.6~MeV. 

The total $^{130}$Te exposure was 19.75 kg$\cdot$yr. The background
at $Q_{\beta\beta}$ was 0.17~cnts/(keV$\cdot$kg$\cdot$yr) if normalized
to the total mass, i.e.~similar to Heidelberg-Moscow. No signal
was found and a lower limit of $T_{1/2}^{0\nu}> 2.8\cdot 10^{24}$~yr
(at 90\% C.L.) was set for $^{130}$Te.  
This limit is not sensitive enough to scrutinize the Heidelberg-Moscow
result.

\subsection{The NEMO-3 experiment}

In NEMO-3 \cite{nemo3}  thin foils made out of 7 
different $\beta\beta$ isotopes (9~kg in total)
were located in a drift chamber with a magnetic field.
% V.I. Tretyak, AIP Conf.Proc. 1417 (2011) 125-128 1.0E24 Mo, 3.2E23 Se
% 
 Outside of
the drift region was a   calorimeter made out of plastic
scintillator blocks with photo multiplier tube (PMT)
readout (FWHM for electrons 15\%/$\sqrt{E[{\mathrm MeV}]}$). 
The $0\nu\beta\beta$ reconstruction efficiency is only about 8\%. 
On the other hand, the topological event
reconstruction largely reduces  backgrounds from locations
other than the source foil as well as internal decays with gammas or
alphas. Only $2\nu\beta\beta$ events with
poor energy reconstruction can not be discriminated. The
 background is about $1.2\cdot10^{-3}$~cnts/(keV$\cdot$kg$\cdot$yr)
at 3~MeV ($\approx Q_{\beta\beta}$ of $^{100}$Mo and $^{82}$Se).
For all isotopes $2\nu\beta\beta$ half lives are reported with
impressive signal to background ratios of up to 76.
For $0\nu\beta\beta$ decay, the 90\% C.L.~limits are 
$T_{1/2}^{0\nu}>1.0\cdot10^{24}$~yr for $^{100}$Mo and 
$T_{1/2}^{0\nu}>3.2\cdot10^{23}$~yr for $^{82}$Se. Again, 
NEMO-3 is not sensitive enough to confirm or reject the Heidelberg-Moscow
claim. 

\subsection{The Kamland-Zen experiment}

In Kamland-Zen \cite{kamland} a balloon of 1.54~m radius made out of
25~$\mu$m thick nylon is inserted into the Kamland detector
and filled with xenon doped scintillator ($\simeq$290 kg $^{136}$Xe). 
The energy (FWHM $\simeq$10\% at $Q_{\beta\beta}$) and
position of the decay 
(resolution $\sigma \simeq 15 \,{\rm cm}/\sqrt{E({\rm MeV})}$)
is reconstructed with  PMTs located
at a radius of 9 m which cover 34\% of the solid angle.

Since the scintillator is very pure and can be doped easily with
several \% xenon, this experiment has the largest target mass
and lowest background - if normalized to the 
total mass.\footnote{The Xe mass fraction
is $\approx$2.5\% and the fiducial volume is $\approx$43\%.} 
However the energy resolution is the poorest.

Kamland-Zen started data taking in 2011. 
An unexpected background peak at about 2.6~MeV is dominating
the spectrum around $Q_{\beta\beta}=2.458$~MeV. It limits the
experimental sensitivity and the best explanation for the
origin is $^{110m}$Ag ($T_{1/2}=250$~d) cosmogenically produced
in $^{136}$Xe while it was above ground. This background
is expected to be reduced to a negligible level by an ongoing
scintillator purification campaign. After an exposure
of 78 days first
results are $T_{1/2}^{0\nu} > 5.7\cdot 10^{24}$~yr (at 90\% C.L.)
and $T_{1/2}^{2\nu} = (2.38\pm0.14)\cdot 10^{21}$~yr \cite{kamland}.
Like for Cuoricino and NEMO-3 this limit is not yet
sensitive enough to scrutinize Heidelberg-Moscow.

Additional 700 kg of Xe with 90\% enrichment are available
by the end of 2012 and are expected to be deployed in a 
cleaner balloon in the near future. Kamland-Zen is therefore
expected to be the first experiment with ton scale isotope mass.
 In the more distant future 
it is foreseen to improve  the energy
reconstruction  with light collectors attached
to the PMTs and a new liquid scintillator \cite{inoue}.

\subsection{The EXO-200 experiment}

EXO-200 \cite{exo} operates a liquid xenon 
TPC of 40 cm diameter and 40 cm length
(175 kg of liquid Xe, 100~kg fiducial mass).
A wire plane in the middle is biased to -8 kV such that electrons drift
to one of the two ends where the ($x,y)$ position of the electron cloud
is reconstructed with 2 planes of wires and the total charge is
measured. The produced scintillation light is detected with 
large area avalanche photo diodes behind the charge collecting wires.

Since there is a strong anti-correlation between the light amplitude
and the ionization signal, combining
both into a new quantity greatly improves
the energy resolution to  FWHM $\approx 3.9$\% at $Q_{\beta\beta}$. The
background  is at a quite low level
of $\simeq 1.5\cdot 10^{-3}$ cnts/(keV$\cdot$kg$\cdot$yr)
 after a geometrical cut on the electron cloud 
is applied to discriminate Compton scattered photon events.

EXO has recently published  results
on  the $2\nu\beta\beta$ half life $T_{1/2}^{2\nu} = (2.1\pm0.2)\cdot 10^{21}$~yr
and  a limit on the
 $0\nu\beta\beta$ half life $T_{1/2}^{0\nu} > 1.6\cdot10^{25}$~yr
(at 90\% C.L.) of $^{136}$Xe \cite{exo}. 

This result can be used to test the claim of Heidelberg-Moscow.
For a given matrix element calculation
the expected number of signal events for EXO can be estimated from
Eq.~\ref{eq:nsig} and \ref{eq:nme11}. 
The  experimental numbers of EXO
 (32.5~kg$\cdot$yr exposure, 55\% reconstruction 
efficiency,\footnote{An efficiency is not explicitly given in
\cite{exo} but an effective value can be estimated using
Eq.~\ref{eq:nsig} and the information from the publication.}
80.6\% enrichment) and Heidelberg-Moscow
(28.75$\pm$6.86 events, 71.7~kg$\cdot$yr exposure, efficiency
100\%, enrichment 86\%) enter. 
EXO-200 reports $n_{\rm obs}=1$ ($n_{\rm obs}=5$) events in
an energy window of $\pm 1\, \sigma$ ($\pm2 \, \sigma$).

A Bayesian formulation can be applied with a null hypothesis ($H$:
EXO observes only background) and
an alternative hypothesis ($\bar{H}$: EXO observes a $0\nu\beta\beta$ signal
with scaled Heidelberg-Moscow event counts). The priors  for the expected
number of signal events $\pi_s$ and for background events 
$\pi_b$ are Gaussian 
and the priors for both hypothesis are set to $\pi(H)=\pi(\bar{H})=0.5$.
Mean  and sigma 
of $\pi_s$ are listed in Tab.~\ref{tab:exo}.
Mean  and sigma 
of the background are  $4.1\pm$0.3 ($7.5\pm0.7$) events
in the  $\pm 1\, \sigma$ ($\pm2 \, \sigma$) energy window.

The posterior probability is then
\begin{equation}
\label{eq:postp}
  p(\bar{H})=
   \frac{ \pi(\bar{H}) \cdot p(D|\bar{H})}
  { \pi(\bar{H}) \cdot p(D|\bar{H}) +
    \pi(H)  \dot p(D|H)}
\end{equation}
with
\begin{eqnarray}
  p(D|\bar{H}) &= &\int  \pi_s(x) 
                         \pi_b(y) 
                          P(n_{\rm obs}|x+y) \,\mathrm{d}x \,\mathrm{d}y \\
  p(D|H) & = & \int   \pi_b(y|\lambda_b,\sigma_b) P(n_{\rm obs}|y)  \,\mathrm{d}y
\end{eqnarray}
Here $P(n_{\rm obs}|y)$ is the Poisson function for mean $y$ and $n_{\rm obs}$ observed events.
The results for $p(\bar{H})$ are listed in Tab.~\ref{tab:exo}.
For the T\"ubingen-Bratislava calculations 
(labelled ``QRPA max'' and ``QRPA min'') 
the maximum and minimum
of 20 ratios $M^{0\nu}(^{136}{\rm Xe})/M^{0\nu}(^{76}{\rm Ge})$ 
are calculated from the values of Tab.~1 of reference \cite{tuebingen12}.
This procedure was suggested by one of the authors \cite{rodinpc}.
For the other models the ratios of
the central values shown in Fig.~\ref{fig:nme} are taken.
None of the results strengthen the hypothesis of a $0\nu\beta\beta$ signal.
The ISM and GCM calculations strongly disfavor them
while the exclusion for the T\"ubingen-Bratislava calculations
are not very strong. The reported EXO-200 data were taken within
a period of 7 months. Hence more stringent statements 
are expected soon.

\begin{table}
\caption{\label{tab:exo} Bayesian posterior probabilities $p(\bar{H})$ 
using EXO-200 data for the
 hypothesis that the $0\nu\beta\beta$ signal of Heidelberg-Moscow
is correct. Probabilities are given for different matrix element
calculations and for the $\pm 1\sigma$ and $\pm 2\sigma$ energy
windows. }
\begin{center}
\begin{tabular}[\columnwidth]{lcccc} \hline 
method     & \multicolumn{2}{c} {in $\pm1 \,\sigma$ window}
           & \multicolumn{2}{c} {in $\pm2 \,\sigma$ window}\\  
           & expected signal evts.  & $p(\bar{H})$ in \% &
                expected signal evts.     & $p(\bar{H})$ in \% \\ \hline
QRPA max & $4.4\pm1.1$ & 4 & $6.1\pm1.5$ & 6  \\ 
QRPA min & $2.8\pm0.7$ & 11 & $3.9\pm0.9$ & 16  \\
ISM & $10.6\pm2.5$ & 0.1 & $14.8\pm3.5$ & 0.2  \\
GCM & $14.3\pm3.4$ & 0.03 & $19.9\pm4.8$ & 0.05 \\
pnQRPA & $6.3\pm1.5 $ & 1 & $8.8\pm2.1$ & 2 \\
IBM & $6.1\pm1.5$ & 1 & $8.6\pm2.1$ & 2 \\ \hline
\end{tabular}
\end{center}
\end{table}

EXO-200 is approved to run for 4 more years.
For a following phase  the spectroscopic identification
of the daughter nucleus of Xe is foreseen. In this case the experiment
will be background free. Only $2\nu\beta\beta$ events with
poorly reconstructed energy can obscure a signal.

\subsection{The GERDA experiment}

 GERDA uses the germanium detectors of Heidelberg-Moscow
and IGEX  and 
- in a second phase - new ones.
The  $\beta\beta$ emitter 
mass is about 13~kg in the first phase.
The detectors are supported by a minimal amount of
material with low radioactivity in a 4~m diameter 
cryostat filled with liquid argon. Argon serves as cooling medium and
shield against external radioactivity. The latter is complemented
by 3~m of water which is instrumented with PMTs to veto
background from muons by the detection
of their Cherenkov light.

GERDA started commissioning in 2010 and found an unexpected large
background from $^{42}$Ar which could be reduced by avoiding
electrical fields around the detectors 
and by an encapsulation of the diodes.
Since November 2011 the first phase of data taking is ongoing.
The background is at the level of 0.02~cnts/(keV$\cdot$kg$\cdot$yr)
and hence almost an order of magnitude smaller than the 
equivalent number of Heidelberg-Moscow.
Due to a data blinding procedure no result on $0\nu\beta\beta$ is expected
before spring 2013. A preliminary result for
$2\nu\beta\beta$ of $T_{1/2}^{2\nu}=(1.88\pm0.10)\cdot 10^{21}$~yr is
reported \cite{gerda}.

A second phase will start early 2013 with additional
new detectors ($\simeq$18 kg of $^{76}$Ge). The background is expected 
to be reduced  from currently 0.02~cnts/(keV$\cdot$kg$\cdot$yr)
to 0.001~cnts/(keV$\cdot$kg$\cdot$yr) due to a liquid argon instrumentation
and a different detector type with enhanced pulse shape discrimination power.

\section{Other future experiments}

The past and running experiments have been discussed above but there
are others under construction or as R\&D efforts.
They use additional isotopes, and various other detection
mechanisms and background reduction methods, see Tab.~\ref{tab:list}.

One important experimental parameter is the 
fraction $\eta$ of the $0\nu\beta\beta$ isotope 
(see Eq.~(\ref{eq:comp}) and Tab.~\ref{tab:g0nu}). $^{130}$Te is the
only one with a large natural abundance. For all other elements
enrichment of the $0\nu\beta\beta$ isotope is mandatory.
$^{136}$Xe is easiest to enrich
with gas centrifuges since this is the heaviest isotope and 
it is already a gas. 
Other materials like $^{76}$Ge can be
converted to a  gas (GeF$_4$) and then
processed. For calcium and neodymium this path
is currently not available and R\&D on alternative methods
is ongoing.

Another important number is $Q_{\beta\beta}$. Larger values are not
only better because of larger $G^{0\nu}$ but also because the
background from natural decay chains falls off fast
beyond 2.6~MeV. Experiments using
$^{76}$Ge have to compensate  by careful material selection and
good energy resolution. 

The experiments can be grouped into 2 classes. In calorimetric
experiments only the total energy (ionization or scintillation) is
measured. In tracking experiments the two electrons are measured
independently, i.e.~the angular distribution between the electrons
is also known. The latter is interesting to study the origin
of the underlying physics in case $0\nu\beta\beta$ is observed.

In the following, the experiments under construction are discussed.

{\bf CUORE} is a continuation
of Cuoricino with close to 1000 TeO$_2$ crystals of 750~g each 
(in total $\simeq 200$~kg of $^{130}$Te). 
The crystal production is almost finished and all major hardware
items are ready or close to. A first tower with 52 crystals (CUORE-0) has
been assembled and data taking in the Cuoricino 
cryostat was scheduled for July 2012 \cite{cuore12}.

After a 
commissioning phase CUORE is expected to start in 2015.

{\bf Majorana} will operate similar germanium diodes 
like GERDA ($\simeq$ 27~kg of $^{76}$Ge) in vacuum
 in a compact 
cryostat made out of electro-formed copper. This self-made
copper is expected to have a factor $>100$ smaller thorium,
uranium and radium contaminations compared to commercial copper
such that the background index is about  0.001~cnts/(keV$\cdot$kg$\cdot$yr).
The shielding is completed by
commercial copper, lead, and polyethylene. 
Operations should start in 2013.

In a later phase a combined GERDA and Majorana germanium experiment with order
ton scale mass is envisioned.

{\bf CANDLES} operates 96 scintillating CaF$_2$ crystals (0.3~kg of $^{48}$Ca)
in a liquid scintillator. Both are in an acrylic container inside a water tank
with PMTs. Data taking started in 2011 but the sensitivity is
limited due to the small target mass. Successful R\&D on $^{48}$Ca enrichment
is crucial for this approach.

{\bf NEXT} is a high pressure xenon gas TPC (1.1~m diameter, 1.4~m length,
pressure 10-15 bar, mass 90-130~kg $^{136}$Xe). 
The  time of the $0\nu\beta\beta$ decay is determined by the detection of the 
(primary) scintillation light with PMTs. The deposited
energy and event topology is
reconstructed from the ionization signal. The drifting electron
cloud passes at the end a volume of higher electric field
such that the electrons are moderately accelerated. 
Consequently, they can excite
xenon but not ionize it. The resulting
(secondary) scintillation light (electro-luminescence) is
 proportional  to the number of electrons and detected with
the same PMTs (for energy reconstruction) and with a plane of 
SiPMs (for position reconstruction). The energy resolution FWHM
is expected to be  $<1$\% at $Q_{\beta\beta}$ and the event
topology will allow to reject backgrounds very effectively
to a level of $8\cdot10^{-4}$~cnts/(keV$\cdot$kg$\cdot$yr).

Construction will start in 2013 and physics data taking in 2015.

{\bf SuperNEMO} is a planned continuation of NEMO-3 with much
improved performance (factor 4 reconstruction efficiency, factor
2 in energy resolution, factor 6 in background).
 A demonstrator module with
7~kg of $^{82}$Se is under construction and expected to start  data
taking in 2014.

{\bf SNO+} dissolves   $\approx$1 ton $^{\rm nat}$Nd in
780 tons of liquid scintillator (44 kg of $^{150}$Nd).
The scintillator is in a 12~m diameter acrylic vessel which is
surrounded by pure water in an 18~m diameter water tank. The latter holds
the 9500 8 inch PMTs for light detection. An energy 
resolution FWHM of $\approx$7\%
at $Q_{\beta\beta}$
is expected with an extremely low background of 
$\le 10^{-6}$~cnts/(keV$\cdot$kg$\cdot$yr) if normalized to the
total scintillator mass. The main background is $2\nu\beta\beta$
due to the poor energy resolution and the relatively short
half life $T_{1/2}^{2\nu}$ of $^{150}$Nd (see Tab.~\ref{tab:g0nu}).
Scintillator filling and doping with Nd is
expected for 2013. 
% bkg = 30 events per bin for 50 keV bin size, 3 years, 780 tons
Since $^{150}$Nd can not be enriched effectively at the moment,
$\eta$ is small which limits the SNO+ sensitivity.

{\bf Lucifer} is an R\&D effort which investigates scintillating
crystals with $\beta\beta$ emitters like ZnSe which are operated
as bolometers. The simultaneous detection of phonons and photons
allows to identify backgrounds from e.g.~surface events which
are expected to dominate in CUORE.

Other ongoing R\&D efforts are not
discussed here. References are listed in Tab.~\ref{tab:list}.

\begin{table}
\caption{ \label{tab:list} Selection of $0\nu\beta\beta$ experiments.}
\begin{centering}
\begin{tabular}[\columnwidth]{lccccc} \hline
experiment & isotope & mass [kg] & method & start / end & ref.  \\ \hline
\multicolumn{6}{c}{past experiments } \\ 
Heidelberg-Moscow & $^{76}$Ge & 11 & ionization & -2003  &\cite{klapdorplb}  \\
Cuoricino & $^{130}$Te & 11 & bolometer & -2008  & \cite{cuoricino} \\
NEMO-3 & $^{100}$Mo, $^{82}$Se & 7,1 & track. +calorim. & -2011 & \cite{nemo3}\\ 
\multicolumn{6}{c}{ current experiments } \\
EXO-200    & $^{136}$Xe & 175 & liquid TPC  & 2011- & \cite{exo} \\
Kamland-Zen & $^{136}$Xe & 330 & liquid scintil.  & 2011- & \cite{kamland} \\
GERDA-I/ GERDA-II   & $^{76}$Ge  & 15/35  & ionization     & 2011-/ 2013- & \cite{gerda} \\
CANDLES     & $^{48}$Ca  & 0.35 & scint. crystal & 2011- & \cite{candles} \\ 
\multicolumn{6}{c}{funded experiments} \\
NEXT        & $^{136}$Xe & 100 & gas TPC  & 2015 & \cite{next} \\
Cuore0/ Cuore & $^{130}$Te & 10/200 & bolometer  & 2012-/ 2015- & \cite{cuore} \\
Majorana Demo. & $^{76}$Ge & 30 & ionization & 2013 & \cite{john} \\
SuperNEMO demo./total  & $^{82}$Se  & 7/100 & track.+calorim.  & 2014-/?? & \cite{supern}\\
SNO+         & $^{150}$Nd & 44 & liquid scint. & 2013 & \cite{sno} \\
\multicolumn{6}{c}{proposal, proto-typing} \\
Cobra       & $^{116}$Cd &       & solid TPC  & &\cite{cobra}  \\
Lucifer     & $^{82}$Se &      & bolom. +scint. & & \cite{lucifer} \\
DCBA/MTD    & $^{150}$Nd & 32 & tracking             & & \cite{dcba} \\
MOON        & $^{82}$Se, $^{100}$Mo & 30-480  & track. +scint.  & & \cite{moon}  \\
AMoRE      & $^{100}$Mo & 100  & bolom. +scint. & & \cite{amore} \\ 
Cd exp.    & $^{116}$Cd &      & scint. & &  \cite{cdexp} \\ \hline 
\end{tabular}
\end{centering}
\end{table}

\section{Comparison of experiments}

For a comparison of the sensitivities of the experiments
 a relative scaling factor for the
different matrix elements and phase spaces has to be applied. This
factor can be estimated from Fig.~\ref{fig:nme2}. The values used 
here are $f_A({\rm Ge})=0.35 (0.2-0.5)$, $f_A({\rm Se})=1.1 (0.7-1.4)$,
$f_A({\rm Mo})=2.1 (1.1-3.2)$, $f_A({\rm Te})=1$,
$f_A({\rm Xe})=0.55 (0.4-0.7)$ and $f_A({\rm Nd})=1.2 (0.4-2.0)$.
The numbers in parentheses are the full range.

If  the number of background events is
large, Eq.~(\ref{eq:comp}) can be used to estimate the experimental
sensitivity. 
A relative figure-of-merit can  then be defined as
\begin{equation}
 {\rm FOM} = f_A\cdot\epsilon\cdot\eta\cdot\sqrt{\frac{M}{B\cdot\Delta E}}
\end{equation}
The relative $T_{1/2}^{0\nu}$ sensitivity scales with
the live time $t$ of an experiment like  FOM$\cdot\sqrt{t}$.
Tab.~\ref{tab:comp} lists the performance numbers of the experiments
discussed above. ``Kamland-Zen2'' is the improved experiment after the
purification of the scintillator (assumed factor 5 smaller total background) and
the upgrade to one ton  xenon mass. For comparison, the FOM numbers,
the expected 90\% C.L.~$T_{1/2}^{0\nu}$ limits for 4~yr of live time, and the
corresponding $m_{\beta\beta}$ limits are given. 
For the latter, the entire spread of the
matrix elements of Fig.~\ref{fig:nme} including the error bars are used.

For running (and past) experiments  the  achieved performance values
are used which might improve with time. For the others the anticipated
performance numbers are taken.

\begin{figure}
\begin{center}
\includegraphics[width=0.7\columnwidth]{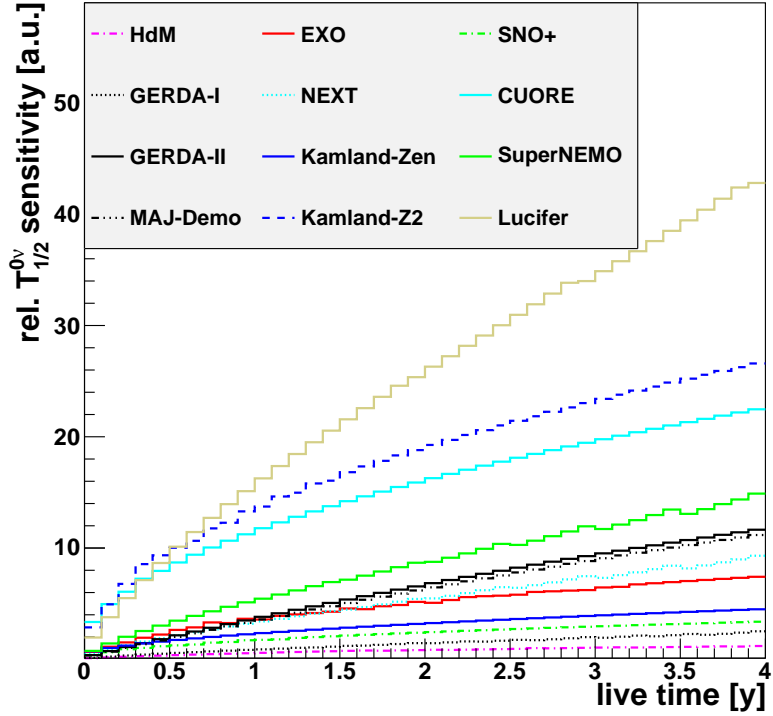}
\end{center}
\caption{\label{fig:t12} Relative experimental sensitivity for the
$0\nu\beta\beta$ half life limit versus live time for
different experiments.}
\end{figure}

As a graphical representation, the relative sensitivity of the
experiments as a function of live time is shown in Fig.~\ref{fig:t12}
This value is calculated from Eq.~(\ref{eq:nsig}) by
\begin{equation}
    \hat{T}_{1/2}^{0\nu} > \frac{f_A \cdot \epsilon \cdot \eta \cdot M \cdot t}
              {\Psi (B\cdot \Delta E \cdot M \cdot t)} 
\end{equation}
Here $\Psi (\lambda_{\rm bkg})$ is the ``average'' 90\% C.L.~upper 
limit of the number of signal
events for $\lambda_{\rm bkg}$ background events calculated according to the
method discussed in \cite{gomez1}. 

\begin{table}
\caption{\label{tab:comp} Comparison of relative figure-of-merit (FOM), lower half life limit $T_{1/2}^{0\nu}$
after 4~yr live time,
and resulting upper limit on $m_{\beta\beta}$.
For $m_{\beta\beta}$, the entire range of matrix
element values including the indicated error bars in Fig.~\ref{fig:nme} are used.
$f_A$ is the average scale factor for
a given isotope taken from Fig.~\ref{fig:nme2}.
$\Delta E$ is the energy window which is taken to be 1(2) FWHM
for experiments with $>0.5\%$ ($<0.5\%$) resolution. Note that the efficiency is
reduced by 0.7 if $\Delta E = $~1$\cdot$FHWM.
FOM is defined in the text. Masses are total masses or fiducial masses. The background
and enrichment fraction has to be scaled accordingly.}
\begin{centering}
\begin{tabular}[1.9\columnwidth]{lccccccccc} \hline
exp. & mass & $f_A$ & bkg. & $\Delta E$ & eff. & enrich. &  FOM & $T_{1/2}^{0\nu}$ & $m_{\beta\beta}$ \\
          &[kg]& & [$\frac{{\rm 10^{-3}cnt}}{{\rm keV}\cdot{\rm kg}\cdot{\rm yr}}$] & [keV] & & & & $10^{25}$~yr & meV\\ \hline 
\multicolumn{10}{c}{past experiments} \\
Hd-Moscow  & 11   & 0.35   & 120      & 7    &  1      & 0.86    & 1  & 1.9 &  170-530 \\
Cuoricino  & 41   &  1     & 170      & 16   &  0.9       & 0.28  &  1  & 0.4 &  210-500 \\
% 11/41=0.27 ENR, FWHM6-12 keV, eff=0.9 geraten
NEMO-3     & 6.9  & 2.1    &  1.2     & 400  &  0.06      & 0.9    & 0.3  & 0.1 & 310-900 \\
% FWHM abgelesen vom plot, eff=8%*0.7, bkg =1.2e-3 vom TAUP vortrag
\multicolumn{10}{c}{running experiments} \\
EXO-200    & 100  & 0.55   & 1.5   & 100   &  0.55      & 0.81   & 6 & 4.2 & 75-170\\
%32.5kgy in 2896h=0.33-> 98kg, sigma=1.67%@2.48->FWHM=100,eff=0.914*0.71*0.825*0.7,bkg=4evt/82keV/32kgy
Kaml.-Zen& 12800 & 0.55   & 0.05    & 250  &  0.31      & 0.023 & 4  & 2.6 & 90-220 \\
% Inoue: 320kg Xe=2,5%->mass SC=12800 kg, 0.9ENR-> enr=0.9*0.025 
% sigma=6.6%/E->FWHM=244,bkg=10evt/112d/50keV/12.8t, eff=fid vol = 0.43,*0.7=0.31
Kaml.-Zen2& 12800 & 0.55   & 0.01    & 250  &  0.31      & 0.06 & 22  & 15 & 40-90 \\
% Inoue: factor 3 more mass, SC cleaning
GERDA-I    &  15  & 0.35   & 20      & 8   &  0.8       & 0.86   & 2 &  3.9 & 120-370 \\ 
GERDA-II   &  35  & 0.35   & 1     &  6   &  0.85      & 0.88   & 20 & 18 & 60-170 \\
\multicolumn{10}{c}{experiments under construction} \\
Major.-Dem.  & 30   & 0.35   & 1   &  6   &  0.9       & 0.9    & 20 & 17 & 60-170\\
CUORE      & 750  & 1      & 10      & 12   &  0.9      & 0.27   & 19 & 7.5 & 50-110 \\
SNO+       & 780000  & 1.5    & 0.0002    & 230  &  0.33     & 5.6E-5  & 3 & 0.8 & 100-240\\
% 30 evt/50keV/3y/780 ton, 0.5 fid vol*0.7 Energy cut, 6.8% FWHM @ 3MeV (TAUP)
NEXT       & 100  & 0.55   & 0.8    & 25   &  0.25      & 0.9    & 9  & 5.2 & 70-160 \\
% NEXT bkg eq 12 in ref JINST,
\multicolumn{10}{c}{proposed experiments} \\
S.NEMO     & 100  & 1.1    & 0.1    & 200  & 0.2        & 0.9    &  14 & 6.9 & 55-140 \\
% 30%eff*0.7, sigma E factor 2.15 < NEMO-3, Bi214 foil bkg factor 10 lower, 2nubb bkg factor 20 lower
Lucifer    & 100  & 1.1    & 1     & 10   &  0.9       & 0.5    &  50 & 19 & 33-85 \\ \hline
\end{tabular}
\end{centering}
\end{table}

A few comments should be made concerning the 
interpretation of Tab.~\ref{tab:comp} and Fig.~\ref{fig:t12}.
\begin{itemize}
\item The factor $f_A$ and hence FOM
      has a full spread (not $\sigma$) of $\approx \pm$ 30\%-70\%.
      Thus the curves in Fig.~\ref{fig:t12} could be replaced by
      bands which would make the figure however unreadable.
\item The sensitivities discussed here are calculated for 
      $0\nu\beta\beta$ exclusion limits.
      For a positive signal claim, the situation is  different.
      A  good energy resolution like the ones for  germanium
      or  bolometer experiments will allow to identify a narrow
      line at the correct energy. 
      This is extremely valuable if the existence of the rarest 
      ever observed decay will be claimed.
\item To estimate relative sensitivities for
      $m_{\beta\beta}$, the inverse square root has to be taken 
      of the curves shown in Fig.~\ref{fig:t12} and the FOM
      numbers in Tab~\ref{tab:comp}.
      The variation due to the spread of the matrix elements is
      reduced in this case as can be seen from the last column
      of Tab.~\ref{tab:comp}, i.e.~the spread for the
      lower or upper value of the $m_{\beta\beta}$ interval is reduced.
      
      If instead limits for other particle physics parameters
      are calculated, the scaling will be different. For
      heavy neutrino exchange, the mass limit scales 
      with $\sqrt{\hat{T}_{1/2}^{0\nu}}$. The  factor
      $f_A$ will change in this case but for most models 
      the variations is less than 30\% \cite{susytub,deppisch}.
\item The $2\nu\beta\beta$ background is irreducible and can only
      be avoided with an energy resolution $\sigma < 1-2$\% at
      $Q_{\beta\beta}$. This requirement depends of course strongly
      on $T_{1/2}^{2\nu}$ which varies by a factor of 300 for
      the isotopes considered. 
\item Of the ongoing experiments, Kamland-Zen(2) should have the
      largest potential. However a low background bolometer
      experiment like Lucifer
      or an improved CUORE experiment will be even more sensitive.
      The sensitivity of them would still grow
      almost linearly after 4~yr (see Fig.~\ref{fig:t12}).
\item Germanium experiments can be  competitive to e.g.~current
      xenon experiments with a factor of 3 more mass despite
      the fact that the phase space factor is  small. 
      Required is however a factor of 10 lower background
      than the one of GERDA-I.
\item Systematic effects like the uncertainty on the fiducial volume
      or the knowledge of the background  level are not taken into 
      account here.
\end{itemize}

The goal of many searches is to reach a sensitivity equivalent
to $m_{\beta\beta}\approx 19$~meV which would cover practically
the entire expected range for the inverted neutrino mass hierarchy.
For $^{76}$Ge, this corresponds to half lives
of $(1.5-15)\cdot 10^{27}$~years if the entire
span of matrix elements of Fig.~\ref{fig:nme} is taken into
account. These values should be compared
to the expected sensitivity of GERDA-II or  Majorana Demonstrator
of about 1.7$\cdot 10^{26}$~y.  This demonstrates that
exploring the entire mass band of the inverted hierarchy is a
long term enterprise.\footnote{To reach a limit of $10^{28}$yr,
a 1 ton $^{76}$Ge experiment has to operate for 10~yr with a background which
is a factor of 10 smaller than the one of GERDA-II.} The equivalent numbers for $^{136}$Xe,
$^{82}$Se, and $^{130}$Te are $(0.6-3.5)\cdot 10^{27}$~yr,  $(0.6-3.8)\cdot 10^{27}$~yr,
and $(0.5-2.5)\cdot 10^{27}$~yr,
respectively. They seem to be easier to reach if one compares
them to the $T_{1/2}^{0\nu}$ limits in Tab.~\ref{tab:comp}.

Considering the example of the Heidelberg-Moscow claim and the
uncertainties of the matrix elements, one can
conclude that a signal has to be observed
in several isotopes to establish $0\nu\beta\beta$.

To improve the credibility of the result a blind analysis
should be performed. This technique is nowadays standard in particle
physics experiments and should be adopted in this field as well.

It is also worth mentioning that continuously new ideas 
for $0\nu\beta\beta$ experiments are coming up \cite{graxe}
%continuously new idea like graxe 1110.6133
and the one who will make a discovery might not be 
listed in this publication.

%beyond to meV scale A.S. Barabash AIP Conf.Proc. 1417 (2011) 5-11

\section{Summary}

Neutrinoless double beta decay  violates
lepton number and the experimental programs are therefore on 
equal footing to proton decay searches.
It might also be the only practical process which
allows to test whether neutrinos are Majorana particles.
The motivation for several large efforts in this field is therefore obvious.

For a long time, the Heidelberg-Moscow experiment has dominated the field and
its claim of a $0\nu\beta\beta$ signal has not been scrutinized since 2001.
The recent EXO-200 limit does not support this claim but can not refute it
due to the spread
of the nuclear matrix element calculations. GERDA does not suffer from
such uncertainties and will unblind the data in spring 2013. 
Then the combined data from GERDA, EXO-200 and Kamland-Zen should
be sensitive enough for a meaningful test.

Beyond this next step, experiments want to reach
a sensitivity to explore the $m_{\beta\beta}$
region of the inverted neutrino mass hierarchy. 
This will eventually require ton scale
experiments. Kamland-Zen will be the first one but whether the entire range
will be covered will depend on the achievable background level.

For a convincing claim of a $0\nu\beta\beta$ signal a good energy resolution is
important and the detection with several isotopes.

\section*{Acknowledgements}

The motivation for this article and many improvements are due
to discussions with several colleagues. I want to thank especially
Allen Caldwell, Karl Tasso K\"opfle, and Stefan Sch\"onert.

\section*{References}
\bibliography{0nubb_review}

\end{document}